# CAPQ-FAST: Content-Adaptive Perceived Quality Assessment for Faster Audiovisual Playback

Jiarun Song, *Member, IEEE*, Yuxin Song, Fuzheng Yang, *Member, IEEE*, Weisi Lin, *Fellow, IEEE*

***Abstract*—Faster playback has become a common feature in modern online audiovisual services, allowing users to consume content in less time while stilling maintaining a coherent viewing experience. However, different modalities of media content, such as video, audio (including speech and music), and audiovisual, exhibit varying requirements for understandability and information integrity under faster playback. These differences lead to noticeable variations in perceived quality depending on the content type. Nevertheless, users' perceived quality at different playback speeds remains insufficiently investigated. To address this gap, this paper conducts a series of subjective experiments to analyze the relationship between playback speed and perceived quality for video, audio, and audiovisual content. Content-specific intrinsic features are extracted to capture temporal dynamics, including temporal information (TI) for video, words per minute (WPM) for speech, and beats per minute (BPM) for music. Predictive models of perceived quality under faster playback are then developed separately for video, speech, and music. By integrating these models, a unified content-adaptive perceived quality assessment model (CAPQ-FAST) is proposed for faster audiovisual playback. Experimental results demonstrate that the proposed model effectively predicts perceived quality under different playback speeds. This model can help service providers better understand users' viewing intentions and perceptual experiences under faster playback, thereby enabling more adaptive personalized recommendations and playback control to enhance user experience adaptability.**

## I. INTRODUCTION

With the rapid growth of short videos, online education, and video-on-demand services, faster playback has become a common user behavior [1]-[6]. Platforms such as YouTube, Netflix, TikTok now support variable playback speeds (e.g., 1.5x, 2x), enabling more efficient information acquisition and allowing users to complete audiovisual tasks within limited time. Beyond improving operational flexibility and personalization, this feature also reflects the user's growing autonomy in media consumption.

This work was supported in part by the National Natural Science Foundation of China (62171353).

Jiarun Song is with the School of Telecommunications Engineering, Xidian University, Xi'an, 710071, China (e-mail: jrsong@xidian.edu.cn).

Yuxin Song is with the Guangzhou Institute of Technology, Xidian University, Guangzhou, 510555, China (e-mail: yxsong24@stu.xidian.edu.cn)

Fuzheng Yang is with the School of Telecommunications Engineering, Xidian University, China, and with the School of Electrical and Computer Engineering, RMIT, Australia (e-mail: fzhyang@mail.xidian.edu.cn).

Weisi Lin is with the College of Computing and Data Science, Nanyang Technological University, Singapore 639798 (e-mail: wslin@ntu.edu.sg).

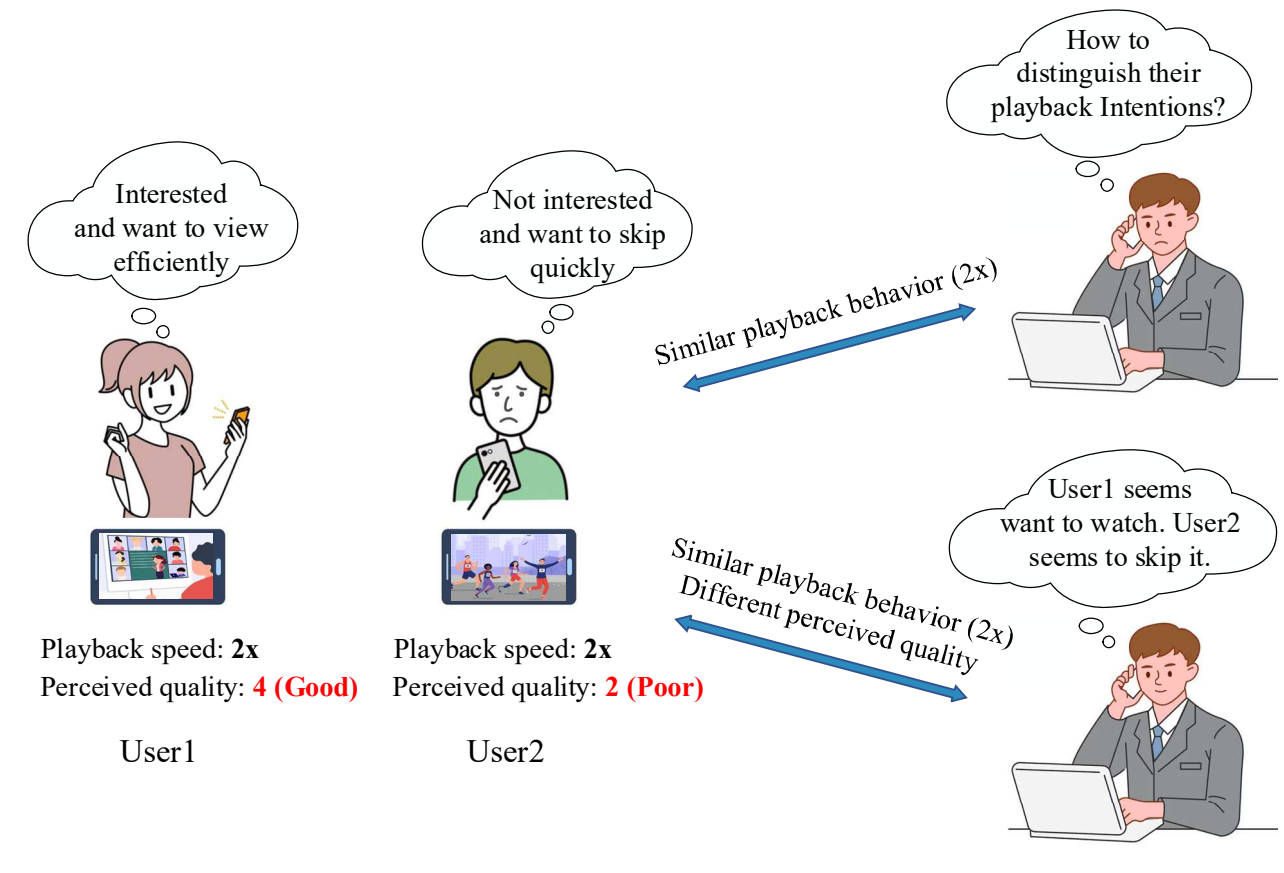

Fig.1. User intention prediction with a same playback speed.

From the perspective of user behavior and cognitive psychology, faster playback essentially reflects a user's active strategy to balance time constraints, cognitive resources, and consumption goals [7], [8]. According to the uses and gratifications theory [9]-[11], media users tend to adopt higher playback speeds when faced with slow-paced content, redundant information, or time pressure, aiming to improve viewing efficiency. However, adjusting playback speed is not always a purely positive behavior. Based on the cognitive load theory (CLT) [12], human information processing capacity and cognitive resources are inherent limits. When information density rises quickly, as in high-speed playback, users may experience cognitive overload, which negatively affects comprehension and perceived quality [13]. This is particularly evident in cases where the content is complex or fast-paced. In such situations, faster playback may not indicate a pursuit of efficiency for but rather reflect a user's attempt to skip, avoid, or disengage from content they perceive as uninteresting or cognitively demanding [14], in search of material that better matches their interests.

Although users can actively adjust playback speed to balance cognitive resources and consumption goals, content service providers often struggle to accurately infer users' viewing intentions and experiences based solely on playback speed behavior. This challenge arises from the inherent ambiguity of faster playback, as the same speed may reflect completely different user intentions depending on the content and context. For example, as illustrated in Fig. 1, both User1 and User2 consume videos at 2x speed, but their intentions differ: User1 accelerates the playback of a slow-paced spoken program (e.g., a lecture) to improve information efficiency, whereas User2 speeds through a fast-paced sports scene (e.g., a competitive chase) to skip irrelevant segments. In this case, relying solely on playback speed poses a significant challenge for the design of recommendation systems and playback strategies on content platforms. This semantic ambiguity of playback speed presents a significant challenge for the design of recommendation systems and playback strategies on content platforms. If the system misinterprets user intent and applies inappropriate quality-reduction measures, such as lowering resolution or bitrate during scenes where users are actually trying to concentrate, it may degrade the user experience. Conversely, failing to optimize resource allocation for segments users intend to skip can lead to bandwidth waste and reduced energy efficiency.

In fact, incorporating perceived quality can help mitigate this issue. For example, in User2's case, if the perceived quality at 2x speed drops significantly and hinders content comprehension, it may suggest that the user is not focused on the current content but is instead attempting to quickly skip through it. In contrast, for User1, if the visual content remains easy to follow and the perceived quality stays high at 2x speed, it indicates that the user is likely aiming to improve viewing efficiency. Therefore, accurately assessing perceived quality under variable playback speeds is of great importance. It provides a valuable basis for improving recommendation strategies and personalized playback control.

However, different modalities of multimedia content, such as video, audio, and audiovisual content, have varying requirements for understandability and information integrity under faster playback. These differences result in significant variations in perceived quality across content types. Our previous research on faster video playback has confirmed this phenomenon [15]. Nevertheless, there is still a lack of systematic studies across multiple modalities, and effective objective evaluation methods have yet to be developed. To address this gap, the present study evaluates perceived quality under faster playback conditions, with particular emphasis on information comprehensibility. Different from general perceptual quality models, we focus on the faster playback scenario and develop a cross-modal, content-adaptive framework to characterize how playback speed and modality-specific temporal content features jointly affect perceived quality. Based on this framework, we propose CAPQ-FAST, a content-adaptive objective perceived quality model for faster audiovisual playback. By providing a perceptual-side cue for understanding faster playback behavior, the model can help service providers optimize personalized playback and recommendation strategies, such as playback speed recommendation, adaptive playback control, and perceptual resource allocation. The main contributions of this paper are summarized as follows:

(1) The perceived quality of different media modalities under faster playback speeds is systematically examined, including video, audio, and audiovisual content, with attention to how unimodal inputs (video and audio) contribute to the audiovisual perception.

(2) For video, speech, and music, content-specific intrinsic features are extracted to characterize modality-specific temporal information density. These features include temporal information (TI) for video, words per minute (WPM) for speech, and beats per minute (BPM) for music, which respectively reflect motion dynamics, linguistic information rate, and rhythmic tempo.

(3) By integrating playback speed with modality-specific content features, a content-adaptive objective model is proposed for perceived quality assessment under faster playback conditions, providing both theoretical support and practical guidance for playback optimization and personalized recommendation.

The rest of this paper is organized as follows. Section II reviews related work. Section III describes the experimental design. Section IV analyzes the relationship between perceived quality and playback speeds and proposes the objective models. Sections V and VI present the performance evaluation and conclusions, respectively.

## II. Related Work

This section provides a brief overview of the two main research areas related to faster playback, namely cognitive load of faster playback and playback optimization. The specific contents are outlined as follows.

### *A. Faster Playback and Cognitive Load*

According to Sweller's cognitive load theory [12], comprehension is significantly impaired when the speed of information presentation exceeds the cognitive processing capacity [16]-[18]. The phonological loop, which plays a key role in auditory information processing, is estimated to have a capacity of approximately 275 words per minute [19]. When playback speed exceeds this processing limit, cognitive resources are rapidly exhausted, which in turn negatively affects learning performance.

Several studies have indicated that when consuming audiovisual content that contains speech (such as the narrator's voice) and visual components (such as text or graphics), faster playback reduces comprehension and memory [20]-[24]. This effect becomes particularly pronounced once playback speed exceeds a certain threshold, typically around twice the normal speed [25]. In other words, when playback speed exceeds the cognitive system's processing capacity, the efficiency of information encoding decreases and comprehension ability declines accordingly. Although prior research has qualitatively identified the impact of faster playback on cognitive performance, there remains a lack of systematic modeling and quantitative analysis of the relationship between playback speed and cognitive outcomes.

In practical applications, research on playback speed is mainly concentrated in the field of education, focusing on its impact on learning efficiency and comprehension. Existing studies suggest that moderately increasing playback speed can enhance time efficiency [26]-[31]. Specifically, as playback speed increases, students are able to complete more learning tasks in less time, thereby demonstrating higher learning efficiency. In addition, several studies have examined how playback speed influences comprehension of media content [32]-[36]. Some studies report that as the playback speed increases, test performance declines, suggesting an upper limit to the amount of information that individuals can process per unit time [32]-[35]. However, other studies indicate that increasing playback speed within a specific threshold does not significantly impair comprehension ability [1], [36]. For instance, [36] found that comprehension was not noticeably affected when audio playback speed did not exceed 1.75x. Similarly, [1] reported that even at 2.5x speed, participants' performance in comprehension tests did not significantly decrease. These findings suggest that users can adapt to faster playback and maintain effective learning and comprehension as long as the speed stays within their cognitive capacity.

Despite extensive prior research on fast playback, it is important to note that most existing studies have primarily focused on the educational domain, where study materials are typically lecture videos that rely heavily on spoken narration. Such videos usually feature relatively simple content structures, consisting mainly of concise text and illustrations, with audio serving as the primary channel of information delivery. However, real-world online video platforms such as YouTube and TikTok provide much more diverse content and media formats [37], incorporating video, speech, background music, and their combinations, while often presenting highly dynamic audiovisual content. In these scenarios, users may accelerate playback not only to improve learning efficiency, but also to browse, screen, or skip short videos, advertisements, movie clips, social media videos, and entertainment content. Therefore, background music, as a common audio component in real-world audiovisual content, may also affect the overall perceived quality and acceptability under faster playback. This further highlights the need to examine faster playback from a cross-modal perspective. Nevertheless, there is still a lack of systematic and unified investigations into perceived quality and user behavior across different media modalities under faster playback conditions.

### B. Faster Playback Optimization

Although faster playback is related to temporal presentation, it should be distinguished from reduced-frame-rate playback. Reduced frame rate mainly introduces temporal discontinuity or motion jerkiness while preserving the original content duration. In contrast, faster playback compresses the temporal presentation of content and increases the amount of visual, auditory, and semantic information delivered per unit time. Therefore, perceived quality degradation under faster playback depends not only on motion smoothness, but also on users' ability to perceive, follow, understand, and accept temporally compressed content.

Extensive research has focused on enabling adaptive playback in video players [38]-[44]. For instance, [38] proposed an intelligent player that dynamically adjusted playback speed based on scene complexity and predefined semantic events. In [39], Zhang et al. introduced the LAPAS algorithm, which combined playback rate and bitrate adaptation to help live streaming service providers explicitly manage latency. Similarly, [40] presented a joint bitrate and playback speed adaptation algorithm to avoid excessively fast playback while keeping latency within the desired range.

However, these adaptive playback technologies primarily focus on system-level performance optimization, such as latency control and bitrate allocation, and often ignore users' perception and underlying intentions when selecting playback speeds. In practice, users' motivations for choosing faster playback are not uniform. Sometimes it is to improve viewing efficiency, and sometimes it is to quickly skip uninteresting content. These differing motivations lead to playback speed changes depending on the context. Therefore, without adequately accounting for users' subjective experiences and behavioral intentions under different playback conditions, adaptive strategies may fail to meet actual user needs and, in some cases, may even degrade the overall experience.

## III. Design of Experiments

To evaluate perceived quality under varied playback speeds, a subjective testing platform was designed for faster playback experiments. Using this platform, a series of subjective tests was conducted on video, audio, and audiovisual content to collect user ratings of perceived quality at different playback speeds. These ratings were used for subsequent perceptual analysis, model construction, and performance validation. The detailed experimental settings including the test platform, participants, and procedures, are described below.

### A. Test Platforms

Currently, most mainstream entertainment and online learning platforms support faster playback. However, users usually need to manually adjust the playback speed each time a new media sequence is played. Such frequent manual speed adjustments may interfere with accurate assessment of user perceived quality when conducting experimental tests on a large number of sequences. To address this limitation, we developed an Android-based subjective testing platform that supports faster playback of video, audio, and audiovisual content. The platform enables playback speeds ranging from 1x to 3x and allows automatic playback based on preset speed settings. Specifically, the platform separately controls video rendering and audio playback. For the video track, Android MediaExtractor and MediaCodec [45] are used for reading, decoding, and rendering. For speeds from 1× to 2×, faster video playback is mainly achieved by scaling the display interval between adjacent frames. For speeds above 2× and up to 3×, fixed frame down-sampling/frame-skipping is further applied together with presentation-time control. For audio playback, speed control is implemented using Android MediaPlayer.PlaybackParams, where the speed parameter is adjusted while the pitch parameter is not additionally modified. Thus, the audio duration changes with speed, but the pitch does not increase proportionally. As shown in Fig. 2(a), when a user selected a test sequence from the list, it played directly at the preset speed, without a manual playback adjustment. After each viewing or listening, a rating interface was shown for users to evaluate quality, as illustrated in Fig. 2(b). The ratings were recorded automatically. All tests were conducted on smartphones with 5.9-inch screens (2560×1440) and headphones.

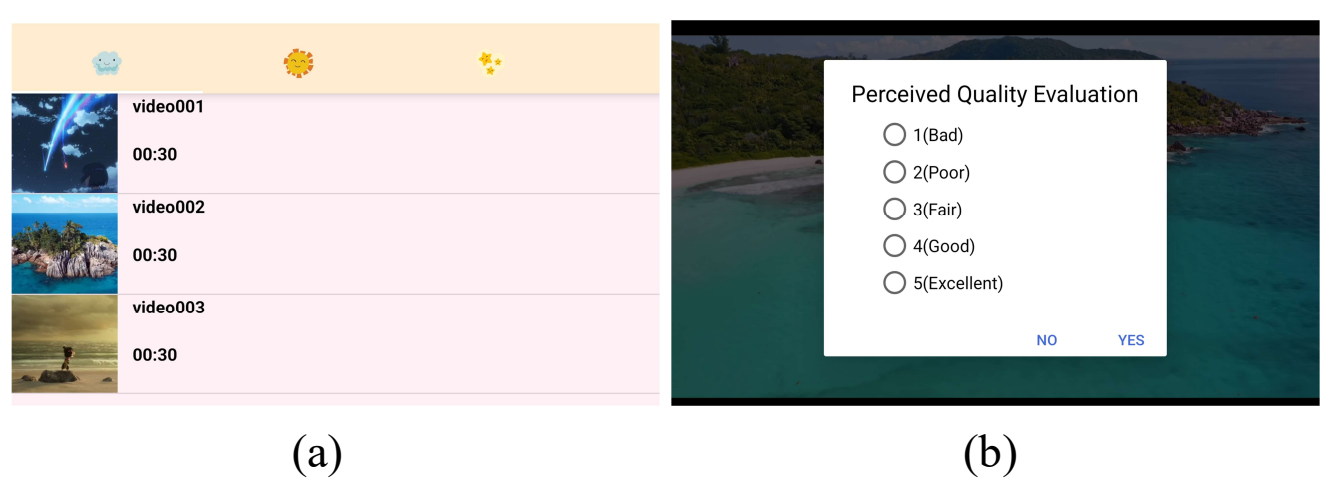

(a) (b)

Fig. 2. Illustration of a test platform. (a) Main interface (b) Rating interface.

*B. Participants*

72 participants took part in the experiment. All were university students between the ages of 22 and 28 (M = 24.2, SD = 2.3), including 33 females (45.8%) and 39 males (54.2%). All participants had normal or corrected-to-normal vision and hearing. Most participants had prior experience with faster playback and were familiar with the faster playback function and its usage context. This helped them provide stable and natural judgments under different playback speeds. However, they typically used faster playback selectively rather than continuously throughout an entire viewing or listening process, which is consistent with realistic user behavior on online media platforms. For each experiment, 24 participants were employed, ensuring that neither gender accounted for less than 1/3 of the group. Participation was entirely voluntary, and informed consent was obtained from all participants prior to the study. Before the formal test began, participants were briefed on the experimental instructions and testing procedures, and they were provided with examples to familiarize themselves with the testing platform, the test flow, and the evaluation criteria. To prevent fatigue, a strict time control strategy was implemented, limiting each testing session to under 20 minutes per participant. It should be noted that the participants were university students, which may limit the diversity of the participant group. However, this group is relevant to faster playback studies because young adults are frequent users of online learning, short-video, and streaming platforms.

*C. Experimental Settings and Procedures*

In this study, perceived quality is defined as users' overall subjective judgment of whether the content can still be clearly perceived, followed, and accepted under faster playback. It involves both perceptual experience and content understanding, with modality-specific emphasis: clarity and understandability for video and speech, and auditory acceptability for music.

Based on this definition, a series of experiments were conducted to investigate perceived quality across different media types and playback speeds. First, a preliminary study was conducted to verify whether the faster playback effect on the test platform we designed is consistent with mainstream platforms. Next, perceived quality evaluation tests were conducted on various media types, including video, audio, and audiovisual content. For video content, the impact of playback speed on perceived quality was examined with respect to different levels of motion complexity. For audio content, speech and music were evaluated separately to explore their respective perceptual characteristics under faster playback. Building on these assessments, further evaluations were carried out for audiovisual content to analyze perceived quality in a multimodal context. The experimental results were used to develop a content-adaptive objective quality assessment model. Finally, the validation experiments were conducted using a broader range of audiovisual content to evaluate the performance of the proposed model.

***Experiment 1***: *Comparison between different players*

In the preliminary test, we compared the playback effect of the developed platform with five mainstream players, namely YouTube, TikTok, Kwai [46], VLC [47] and KMPlayer [48]. The media types compared include audio, video, and audiovisual. Each media modality contains three test sequences, comparing the differences in perceived quality at 1x, 1.25x, 1.5x, 2x, and 3x speeds. Before the test, participants were informed of the experimental procedures and instructions to ensure familiarity with the platform interface, testing process, and evaluation criteria. The experiment adopted the pair comparison method. Each participant selected the same playback speed on the developed and comparison players to view or listen the same test source, and rated the playback difference between players based on their subjective experience, where a score of "1" indicated a noticeable difference, and "0" indicated no significant difference. The order of video, audio, and audiovisual test was random, and the order of test sequences in each media modality was also random. A five-minute break was given between media types, and the process continued until all tests were completed.

***Experiment 2***: *Experiment for Model Training*

This experiment includes three sessions for the purpose of model training. The first session focused on evaluating perceived quality of video content at various playback speeds. The second session targeted the perceived quality of audio content under faster playback. Building upon the first two, the third session investigated perceived quality in the context of audiovisual content played at different speeds.

*Session1: Faster Playback for Video*

In alignment with typical user playback habits, five playback speed levels were included in this experiment, namely 1x, 1.25x, 1.5x, 2x, and 3x. A total of 75 high-quality video sequences (1080p, lossless compression) with different motion complexities were selected. These sequences covered diverse content categories, such as news, movies, dance, sports, and entertainment videos, to reflect different motion patterns and viewing contexts. Based on their Temporal Information (TI) values [49], the sequences were divided into five groups, as shown in Table I. The grouping thresholds were empirically determined based on the TI distribution of the current dataset to obtain gradually increasing levels of motion complexity. From Group1 to Group5, the average TI values gradually increased, reflecting increasing motion complexity. Each group of video sequences was played in a random order of five playback speeds, with three test sequences for each speed.

TABLE I
EXPERIMENTAL SETTINGS OF VIDEO SEQUENCES

| Groups | TI values (Avg) | Num of sequences |
|---|---|---|
| Group1 | 3.19-5.18 (4.42) | 15 |
| Group2 | 8.84-10.35 (9.55) | 15 |
| Group3 | 15.22-17.07 (16.49) | 15 |
| Group4 | 20.39-21.69 (20.65) | 15 |
| Group5 | 26.28-27.85 (27.05) | 15 |

TABLE II
PERCEIVED QUALITY DESCRIPTION FOR VIDEO PLAYBACK

| Label | Scale | Description |
|---|---|---|
| Excellent | 5 | Very easy to see and understand the content |
| Good | 4 | Able to see and understand video content |
| Fair | 3 | Slightly unclear video content, but does not affect understanding |
| Poor | 2 | Somewhat difficult to see the video content clearly and understand the video content |
| Bad | 1 | Unable to see the video content clearly, nor understand it |

During the test, participants selected video sequences from a playlist and viewed them at randomly assigned playback speeds. After the video was played, participants were asked to evaluate the perceived quality based on the absolute category rating (ACR) using a five-point rating scale [50], where the categories “excellent”, “good”, “fair”, “poor” and “bad” corresponds to scores from 5 to 1 respectively. The detailed description of the quality levels is provided in Table II. After the evaluation was completed, the platform automatically returned to the video playlist. To prevent the potential influence from the playback speed of the previously viewed content as much as possible, participants need to wait 3 seconds before selecting the next sequence for testing. It should be noted that participants were not allowed to exit, pause, or interrupt playback during the controlled subjective quality evaluation tests, so that all ratings were based on the same playback speed, complete stimulus exposure, and consistent temporal context. In order to minimize the deviation of the perceived quality, four randomized playback sequences were predefined. Each participant was randomly assigned one sequences, and the mean opinion score (MOS) was used to evaluate perceived quality at different playback speeds.

*Session2: Faster Playback for Audio*

To account for potential differences in user perception of faster playback between speech and music, the audio subjective test was designed to separately evaluate the effects of playback speed on perceived quality for each content type. Specifically, the playback speed was set to five levels, namely 1x, 1.25x, 1.5x, 2x, and 3x. For the speech test, four groups of speech sequences with different speaking rates were selected, covering different spoken-content scenarios such as narration, lecture-style speech, news-style speech, and conversational speech. The speaking rates were measured in terms of words per minute (WPM), and each group included three speech sequences with similar speaking rates. For the music test, four groups of music sequences with different tempos were selected, covering different music styles and rhythmic characteristics. The tempos were measured in beats per minute (BPM), and each group included three music sequences with similar tempos. The settings of speech and music are shown in the following Table III. Each group contains 3 test sequences, and each sequence is played at 5 different playback speeds in a randomized order.

Similarly, during the audio test, each participant was first randomly assigned to begin with either the speech or music condition. Participants selected sequences from the corresponding playlist and listened to them at randomly assigned playback speeds. After each sequence ended, they were required to evaluate the perceived quality using the ACR method on a five-point scale, where the categories “excellent”, “good”, “fair”, “poor” and “bad” corresponds to scores from 5 to 1 respectively. The detailed description of the quality levels is provided in Table IV. When participants completed the evaluation, the test platform automatically returned to the playlist interface. They also need to wait 3 seconds before selecting the next sequence for testing. Similarly, participants were not allowed to exit, pause, or interrupt audio playback during the controlled listening test, for the same reason of maintaining consistent stimulus exposure and reliable MOS statistics. To minimize potential perceptual bias, four randomized playback sequences were predefined. Each participant selected one of these random orders to complete the test. After completing all sequences in the first audio type, the participant proceeded to the other type (speech or music) and repeated the same procedure until all sequences had been evaluated. Finally, the perceived quality at different playback speeds was measured based on the MOS.

*Session3: Faster Playback for Audiovisual*

For the audiovisual subjective test, 20 speech-video sequences featuring various speaking rates and 20 music-video sequences with different tempos were selected, covering a range of content scenarios with different speaking styles, background music characteristics, and viewing contexts. Each sequence was randomly presented at one of five playback speeds, namely 1x, 1.25x, 1.5x, 2x, and 3x. Participants were asked to evaluate the perceived quality using a five-point rating scale. The five rating levels corresponded to the following: “excellent”, “good”, “fair”, “poor” and “bad” mapped to scores from 5 to 1, respectively. The testing procedure followed the same protocol as the other experimental sessions.

***Experiment 3**: Experiment for Validation*

Two validation sessions were designed in this study. The first one involved passive viewing, where participants watched videos at different playback speeds without their control. This session was primarily used to evaluate the predictive performance of the proposed model. The second one involved active playback, where participants freely selected their preferred playback speeds. This setting was used to assess whether the proposed model could more accurately predict users' subjective viewing intentions based on their playback speed behavior.

TABLE III
EXPERIMENTAL SETTINGS OF AUDIO SEQUENCES

| Types | Groups | WPM values (Avg) | Num of seqs |
|---|---|---|---|
| Speech | Group1 | 161-173 (168) | 3 |
| | Group2 | 256-269 (261) | 3 |
| | Group3 | 323-336 (331) | 3 |
| | Group4 | 394-403 (398) | 3 |
| **Types** | **Groups** | **BPM values (Avg)** | **Num of seqs** |
| Music | Group1 | 91-102 (95) | 3 |
| | Group2 | 117-129 (122) | 3 |
| | Group3 | 144-156 (149) | 3 |
| | Group4 | 169-181 (175) | 3 |

TABLE IV
PERCEIVED QUALITY DESCRIPTION FOR AUDIO PLAYBACK

| Types | Label | Scale | Description |
|---|---|---|---|
| Speech | Excellent | 5 | Very easy to listen to and understand the content |
| | Good | 4 | Can listen to and understand content |
| | Fair | 3 | Slightly unclear speech content, but does not affect understanding |
| | Poor | 2 | Somewhat difficult to hear the content clearly and understand the content |
| | Bad | 1 | Unable to hear the speech content clearly, nor understand it |
| **Types** | **Label** | **Scale** | **Description** |
| Music | Excellent | 5 | Completely acceptable |
| | Good | 4 | Fairly acceptable |
| | Fair | 3 | Neutral |
| | Poor | 2 | Relatively unacceptable |
| | Bad | 1 | Completely unacceptable |

*Session1: Experiment for Model Performance Validation*

This experiment was primarily conducted to validate the performance of the proposed model. It involved the subjective evaluation of perceived quality for video, audio, and audiovisual content at different playback speeds. Specifically, for the video playback test, 60 high-quality video sequences were selected (different from those in the training experiment), including scenes with fast, moderate, and slow motion. The TI values of these sequences ranged from 1.21 to 57.58. Five playback speed levels were used, namely 1x, 1.25x, 1.5x, 2x, and 3x. For each speed level, 12 test sequences were selected, with four sequences for high, medium, and low motion complexity category respectively.

In the audio playback test, 20 speech sequences with varying speaking rates (WPM ranging from 164 to 431) and 20 music sequences with different tempos (BPM ranging from 91 to 179) were used. Each sequence was randomly played at one of the five speed levels (1x, 1.25x, 1.5x, 2x, and 3x), and subjective quality ratings were collected. For each playback speed, 4 speech and 4 music sequences were evaluated. For the audiovisual playback test, 35 speech-video sequences and 35 music-video sequences were selected. Each sequence was randomly played at one of the five playback speeds (1x, 1.25x, 1.5x, 2x, and 3x), and subjective quality scores were recorded. For each speed level, 7 speech-video and 7 music-video sequences were evaluated. The experimental procedure was consistent with that of the training experiments. Particularly, all participants in this experiment had not taken part in any of the previous experiments.

*Session2: Experiment for Playback Intention Prediction*

This experiment was designed to examine whether perceived quality can serve as an auxiliary cue for interpreting users' faster playback intentions. Different from the controlled subjective quality evaluation tests, this session aimed to simulate active playback behavior under more realistic viewing conditions. To this end, 56 high-quality audiovisual sequences with a duration of 1 to 2 minutes were selected for free viewing. These sequences covered a variety of real-world content types, including news, movie clips, sports, advertisements, short videos, and entertainment content, to reflect diverse viewing contexts on practical audiovisual platforms. Participants were allowed to choose their preferred playback speed based on their personal viewing preferences for each sequence. If a participant was not interested in a particular sequence, they could skip it entirely, and the corresponding data was excluded from the analysis.

During playback, each time a participant adjusted the playback speed, the system automatically recorded the selected speed and the corresponding time points to facilitate the calculation of objective audiovisual parameters. At the same time, participants were required to indicate the reason for each speed adjustment by selecting one of two predefined options. These two categories were introduced and explained to participants before the experiment. Specifically, Efficiency Viewing (EV) indicated that the participant was interested in the content but wanted to improve viewing efficiency through faster playback, while Selective Skipping (SS) indicated that the participant had lower interest in the current content and wanted to skip through it quickly. This two-option design was adopted to facilitate statistical analysis and establish a clear correspondence between playback intention and perceived quality. The remaining experimental procedure was consistent with that of the previous sessions.

## IV. EXPERIMENTAL RESULTS FOR MODELING

This section first compares playback performance between the developed platform and commercial players. Based on this, objective quality assessment models are constructed for video and audio under varying playback speeds, using their respective content features, namely TI for video, WPM for speech, and BPM for music. Finally, the contributions of audio and video to overall audiovisual quality are analyzed, leading to the CAPQ-FAST model for faster playback.

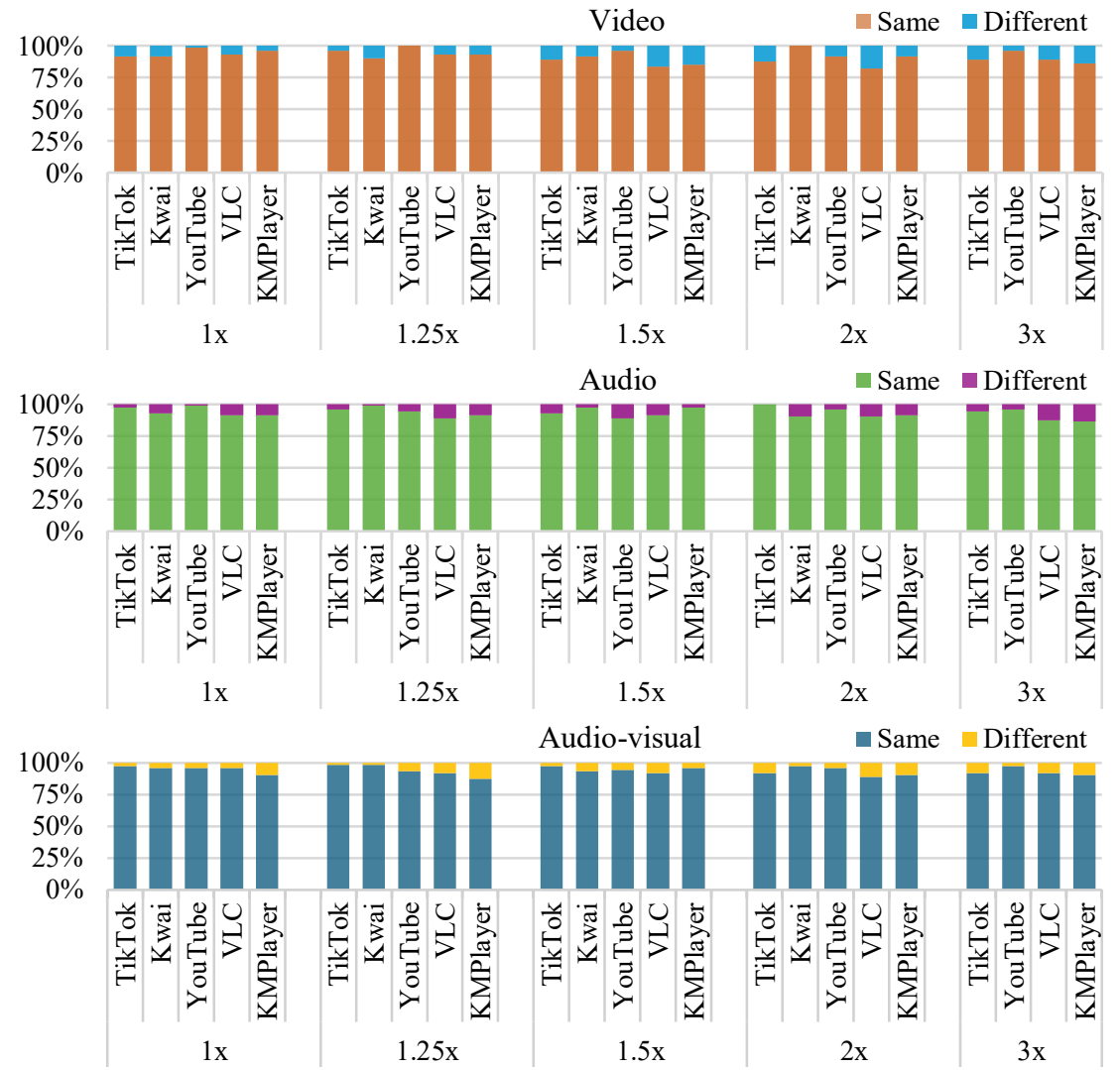

Fig. 3. Fast playback comparison of media players for video, audio, and audiovisual.

### A. Fast Playback Comparison of Media Players

First, a frequency analysis of all pairwise comparison results was conducted, and the stacked bar charts were used for visualization to illustrate the proportion of perceived consistency and difference by different players in various playback speeds conditions, as shown in Fig. 3. The results show that for video playback, the proposed player exhibits minimal differences compared to mainstream players, with an average perceived similarity rate exceeding 88%. Among them, YouTube showed the highest similarity (96.39%), followed by Kwai (93.40%). TikTok, KMPlayer, and VLC performed slightly lower, with similarity rates of 90.56%, 90.28%, and 88.06%, respectively.

For audio playback, the proposed player also demonstrated high consistency with other mainstream players, with average similarity rates above 90%. In particular, TikTok, Kwai, and YouTube all showed similarity rates above 95%, while VLC and KMPlayer were slightly lower, around 90%. Regarding audiovisual playback, the proposed player achieved the highest perceived similarity with TikTok, Kwai, and YouTube at all playback speeds, with similarity rates above 95%. The similarity with VLC and KMPlayer also remained above 90% overall.

In order to evaluate whether different playback speeds have a significant impact on the perceived consistency judgments of each player, chi-square tests were conducted for video, audio, and audiovisual playback. Specifically, we analyzed the frequency of "0" and "1" responses given by users when comparing the proposed player with each target player at different playback speeds. The results show that in video playback, the $\chi^2$ values ranged from 3.703 to 7.553, with p-values between 0.074 and 0.448, all exceeding the 0.05 significance threshold. In audio playback, the $\chi^2$ values ranged from 1.049 to 7.446 and p-values from 0.094 to 0.902, which are also greater than 0.05. Similarly, for audiovisual playback, the $\chi^2$ values ranged from 1.605 to 7.162, with p-values between 0.128 and 0.808. These results indicate that the distribution of perceived consistency under playback speeds was not significantly different ($p > 0.05$), suggesting that playback speed had minimal impact on users' perceived consistency for video, audio, and audiovisual content.

Overall, under the current experimental conditions, the subjective experience difference between the designed player and the target player in the fast playback situation was small. This provides empirical support and technical reliability for conducting further experiments on this platform and enhances the universality of the conclusions.

### B. Perceived Quality Analysis and Modeling

Based on the subjective ratings collected in Experiment 2, we analyzed the variations in perceived quality across video, audio, and audiovisual content under different playback speeds and developed objective assessment models accordingly. In this modeling process, TI, WPM, and BPM are adopted as the primary content descriptors because they provide simple and interpretable measures of modality-specific temporal information density for video, speech, and music, respectively. As playback speed increases, these temporal-density characteristics become closely associated with users' ability to perceive, follow, understand, or accept the content.

#### (1) Perceived Quality Assessment for Faster Video playback

Fig. 4 shows the relationship between perceived quality and playback speed across video groups. It can be found that perceived quality (in terms of MOS) generally decreases as playback speed increases, but the decline trend varies among groups. A Friedman test ($\chi^2$=16.00, p=0.003) revealed significant differences in perceived quality among the five video groups, confirming that playback speed affects different video types to varying degrees, with some groups showing steeper declines than others.

For groups with slow motion characteristics (e.g., Group 1), the decline in perceived quality with playback speed is relatively slow. Compared with normal playback speed (1x), the perceived quality decreases by about 0.69 at 2x speed and about 0.96 at 3x speed. These findings suggest that when the motion characteristics (in terms of TI) of the scene are low, users tend to maintain a high level of perceived quality (scores between 4 and 5). Even at 3x playback speed, most users can watch and understand the video content normally. In this case, playback speed alone is not a reliable indicator of users' willingness to engage with the content.

In contrast, groups with more intense motion scenes, such as Group 5 (Avg TI=27.05), show a rapid decline in perceived quality as playback speed increases. Compared with 1x speed playback, the perceived quality drops by about 2.49 points at 2x speed and by about 3.33 points at 3x speed. When playback speed stays below 1.5x, users generally report high quality (scores between 3 and 5). However, if the playback speed reaches 3x, user's perceived quality will drop to a low level (scores between 1 and 2). In summary, videos with different scene features may yield varying levels of perceived quality even at the same playback speed. Therefore, perceived video quality assessment should consider not only the impact of playback speed but also the motion characteristics of the scene. It should be noted that TI is adopted in this study to characterize the temporal dynamics and motion complexity that directly affect visual comprehensibility under faster playback. It is not intended to fully represent semantic importance, narrative relevance, user interest, or emotional value. These semantic-level factors may further influence users' playback behavior and will be considered in future extensions of the proposed model.

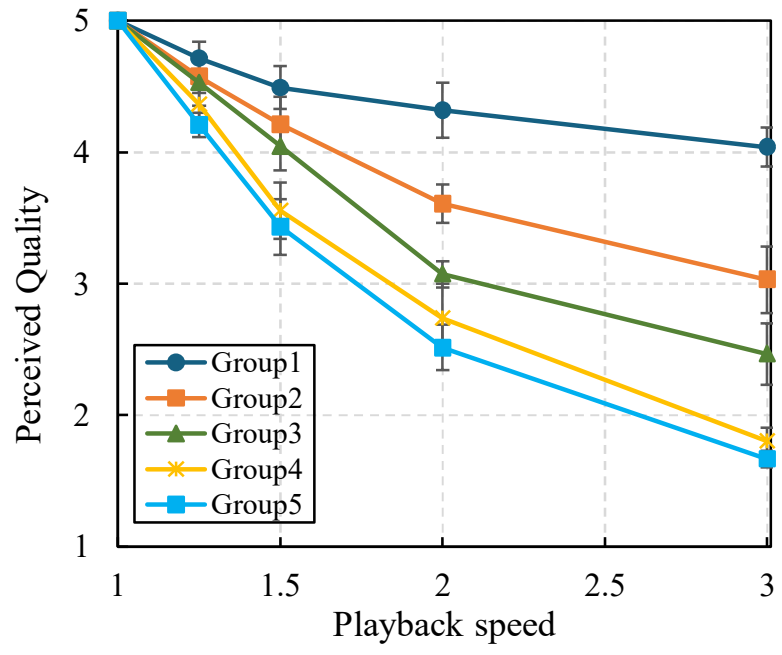

Fig. 4 Relationship between playback speed and perceived quality for video.

The results shown in Fig. 4 indicate that the relationship between perceived quality and playback speed generally follows an exponential degradation trend. Similar exponential forms have been widely used in perceptual quality and QoE modeling. Here, this form is adopted as a compact parametric representation to describe the observed perceived quality degradation under faster playback. Therefore, the video quality model can be expressed as:

$$Q_v = \alpha \cdot \exp(-\beta \cdot S) + \delta \tag{1}$$

where $Q_v$ represents the perceived quality of video under faster playback, while $S$ denotes the playback speed. The parameter $\alpha$, $\beta$, and $\delta$ are estimated using least squares fitting method based on the data presented in Fig. 4. Table V lists the values of these parameters for different video groups. The results show that the differences in $\beta$ under video groups are relatively small (Standard deviation, $SD_\beta = 0.006$). Therefore, $\beta$ can be treated as a constant and is set to the average value of all test video clips, representing the general sensitivity of perceived quality to playback speed. In contrast, the values of $\alpha$ and $\delta$ vary significantly among different groups ($SD_\alpha = 2.639$, $SD_\delta = 1.093$). Fig. 5 illustrates the relationships between $\alpha$, $\delta$, and the TI values of the video in different groups. It can be observed that $\alpha$ increases with higher TI values, whereas $\delta$ decreases as TI increases, indicating that these two parameters capture the content-dependent degradation range and baseline quality level. This trend suggests that videos with higher motion complexity suffer stronger perceived quality degradation under faster playback. Accordingly, $\alpha$ and $\delta$ are treated as content-adaptive parameters and expressed as TI-based functions to capture the

TABLE V
VALUES OF $\alpha$, $\beta$, AND $\delta$

| Groups | TI | $\alpha$ | $\beta$ | $\delta$ |
|---|---|---|---|---|
| Group1 | 4.43 | 2.676 | 0.857 | 3.818 |
| Group2 | 9.55 | 5.692 | 0.849 | 3.589 |
| Group3 | 16.49 | 8.338 | 0.861 | 1.627 |
| Group4 | 20.65 | 9.365 | 0.866 | 1.089 |
| Group5 | 27.05 | 9.674 | 0.854 | 0.839 |

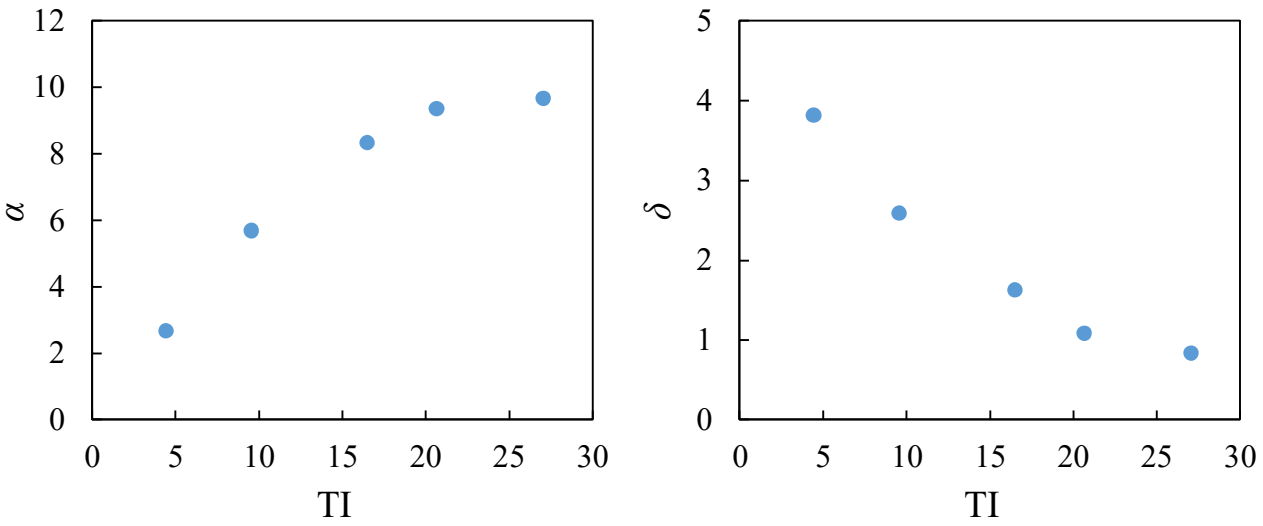

Fig. 5. Relationship between TI and the values of $\alpha$ and $\delta$.

effect of video motion complexity under faster playback. They can be calculated as follows:

$$\alpha = v_1 - \frac{v_1}{\left(1 + TI/v_2\right)^{v_3}} \tag{2}$$

$$\delta = v_4 \cdot \exp\left(v_5 \cdot TI\right) \tag{3}$$

where the parameters in each equation are determined as follows: $\beta = 0.86$, $v_1 = 11.58$, $v_2 = 9.409$, $v_3 = 1.655$, $v_4 = 5.222$, and $v_5 = -0.072$. These values are derived from the data presented in Fig. 5 using least squares fitting method.

*(2) Perceived Quality Assessment for Faster Audio playback*

The following analysis examines how perceived quality varies with playback speed for two common audio formats, namely speech and music. For speech, Fig. 6(a) presents the relationship between perceived quality and playback speed for different groups of speech clips. It can be seen that perceived quality generally declines as playback speed increases, though the rate of decline differs among groups. A Friedman test ($\chi^2 = 12.00$, $p = 0.007$) revealed significant differences among the groups, indicating that playback speed affects the perceived quality of different speech content to varying degrees.

As shown in Fig. 6(a), for speech sequences with slower speaking rates (i.e., lower WPM), such as Group 1, the decline in perceived quality is relatively gradual as playback speed increases. Compared to the normal playback speed (1x), the perceived quality drops by approximately 0.54 at 2x speed and by about 0.71 at 3x speed. This result indicates that when the WPM is low, users generally perceive fast-played speech as maintaining a relatively high-quality level (4-5 points). Even at 3x speed, most users are still able to comprehend the speech content. Therefore, in such cases, playback speed alone is not a reliable indicator of users' willingness to listen.

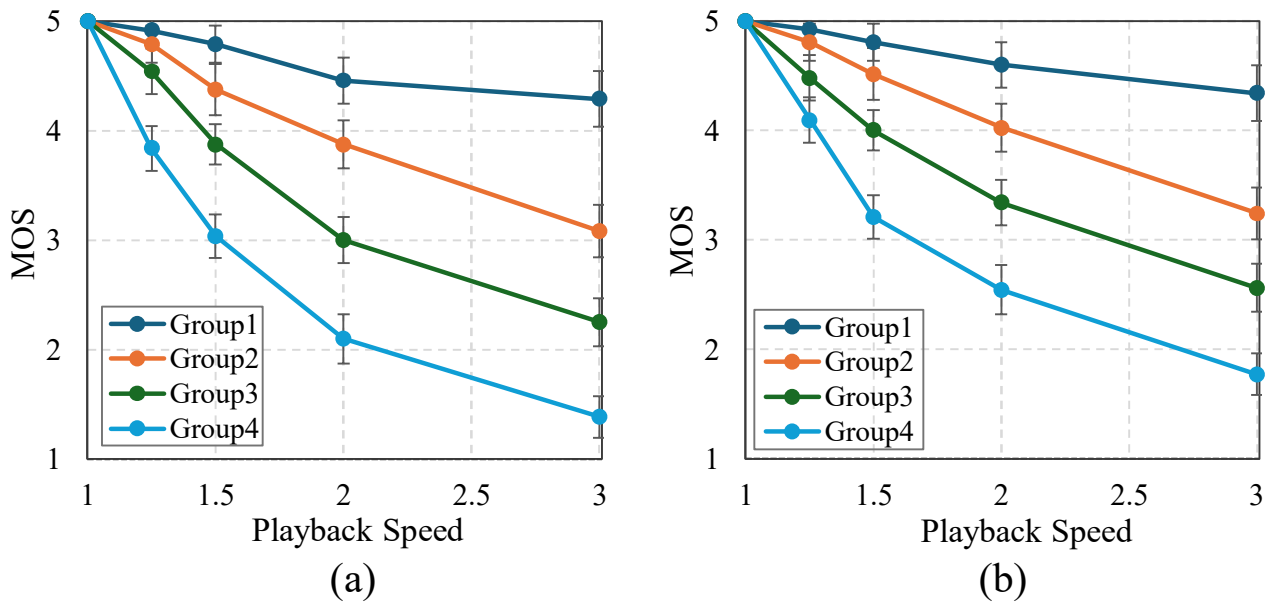

Fig. 6 Relationship between the perceived quality and playback speed for speech and music. (a) Speech, (b) Music.

TABLE VI
VALUES OF $\lambda$, $\mu$, $\rho$, AND $\gamma$

| Groups | *Speech* | | | *Music* | | |
|---|---|---|---|---|---|---|
| | **WPM** | **λ** | ***μ*** | **BPM** | ***ρ*** | ***γ*** |
| Group1 | 168 | 5.718 | 0.112 | 95 | 5.609 | 0.095 |
| Group2 | 261 | 5.714 | 0.183 | 122 | 5.603 | 0.159 |
| Group3 | 331 | 5.716 | 0.264 | 149 | 5.610 | 0.230 |
| Group4 | 398 | 5.713 | 0.427 | 175 | 5.613 | 0.362 |

For speech sequences with faster speaking rates (i.e., higher WPM), such as Group 4, the perceived quality declines rapidly with increasing playback speed. Compared to 1x playback speed, the perceived quality drops by approximately 2.9 at 2x speed and by approximately 3.62 at 3x speed. Overall, when playback speed does not exceed 1.5x, users tend to perceive fast-played speech as acceptable, with quality ratings generally ranging from 3 to 5. However, once playback exceeds 2x speed, perceived speech quality drops to a low level (1-2 points), suggesting that it is influenced not only by playback speed but also by the original speaking rate.

According to the results in Fig. 6(a), the relationship between perceived quality and playback speed for the speech can be expressed by an exponential function:

$$Q_{as} = \lambda \cdot \exp(-\mu \cdot S) \tag{4}$$

where $Q_{as}$ is the perceived quality of the speech, and $S$ is the playback speed. $\lambda$ and $\mu$ are parameters obtained by least squares fitting based on the data presented in Fig. 6(a). Table VI lists the values of $\lambda$ and $\mu$ for different speech sequences. It can be seen that the differences in $\lambda$ values between different speech sequences are small ($SD_\lambda$ = 0.0005). Therefore, the parameter $\lambda$ can be treated as a constant and set to the average value ($\lambda$ = 5.715). In contrast, $\mu$ varies substantially across groups ($SD_\mu$ = 0.102) and generally rises with WPM. This indicates that $\mu$ controls the degradation rate of perceived speech quality with increasing playback speed. A larger $\mu$ corresponds to faster quality degradation, meaning that speech with higher WPM is less tolerant to faster playback. Therefore, $\mu$ is regarded as a content-adaptive parameter and calculated from WPM values using an exponential function as follows:

$$\mu = v_6 \cdot \exp(v_7 \cdot WPM) \tag{5}$$

The parameters $v_6$ and $v_7$, obtained by least-squares fitting from Table VI, are 0.036 and 0.006, respectively.

In contrast to speech, the effects of playback speed on music perception are illustrated in Fig. 6(b), which shows the relationship between playback speed and the perceived quality of different music sequences. The overall trend is broadly similar to that observed for speech, with perceived quality generally declining as playback speed increases. A Friedman test revealed significant differences in perceived quality among the four music groups across playback speeds ($\chi^2$ = 12.00, p = 0.007), indicating that playback speed affects different music content to varying degrees.

For music sequences with slower tempos (i.e., lower BPM), such as Group 1, the decline in perceived quality with increasing playback speed is relatively slowly. Compared to the original speed, the perceived quality decreases by approximately 0.4 at 2x speed and by around 0.66 at 3x speed. This indicates that users generally perceive fast-played music with slower tempos as maintaining a relatively high quality level (4-5 points). Therefore, in such cases, playback speed alone is also not a reliable indicator of users' willingness to listen. By comparison, for music sequences with faster tempos (e.g., Group 4), perceived quality drops sharply as playback speed increases. Compared to 1x speed, quality ratings decrease by approximately 2.45 at 2x speed and by about 3.23 at 3x speed. Therefore, the perceived quality of audio is influenced not only by playback speed but also closely related to the tempo of the music. According to the results in Fig. 6(b), the relationship between perceived quality and playback speed for the music can be expressed as follows:

$$Q_{am} = \rho \cdot \exp(-\gamma \cdot S) \tag{6}$$

where $Q_{am}$ is the perceived quality of the music, and $S$ is the playback speed. $\rho$ and $\gamma$ are parameters obtained by least squares fitting based on the data presented in Fig. 6(b). Table VI lists the values of $\rho$ and $\gamma$ for different music sequences. It can be seen that the differences in $\rho$ values between different music sequences are small ($SD_\rho$ =0.002). Therefore, the parameter $\rho$ can be treated as a constant and set to the average value of the test music sequences. In contrast, the differences in $\gamma$ values between different music sequences are substantially variable ($SD_\gamma$ = 0.06). It can be seen that $\gamma$ values increase with increasing BPM values. This indicates that $\gamma$ controls the degradation rate of perceived music quality with increasing playback speed. A larger $\gamma$ corresponds to faster quality degradation, suggesting that music with higher BPM is less tolerant to faster playback. Therefore, $\gamma$ is regarded as a content-adaptive parameter and can be calculated based on BPM as follows:

$$\gamma = v_8 \cdot \exp(v_9 \cdot BPM) \tag{7}$$

where the parameters are $\rho$ = 5.608, $v_6$ = 0.023, and $v_7$ = 0.016, which are obtained by least squares fitting based on the data presented in Table VI.

*(3) Perceived Quality Assessment for Faster Audiovisual*

The perceived quality of audiovisual content under variable playback speed is a result of users' integrated perception of both accelerated audio and video. Correlation analysis of the experimental data reveals that overall audiovisual quality is significantly associated with video quality, audio quality, and their interaction. Specifically, for speech-video sequences under fast playback, the correlation coefficients between overall perceived quality and video quality, audio (speech) quality, and their coupling are 0.860, 0.934, and 0.926 respectively, all with $p$ values below 0.05, indicating significant relationships in all three cases. Similarly, for fast-playing music-video sequences, the overall perceived quality correlates with video quality, audio (music) quality, and their coupling at 0.972, 0.900, and 0.960 respectively, also with $p$ values below 0.05, confirming statistically significant correlations. Inspired by prior parametric audiovisual quality assessment models, we adopt an audio-video interaction form to combine the predicted video and audio qualities. Different from general audio-visual quality assessment models [51], [52], the inputs $Q_v$ and $Q_a$ here are specifically derived from faster-playback quality models with content-adaptive temporal descriptors. In this regard, a comprehensive audiovisual quality assessment model (CAPQ-FAST) under faster playback conditions is constructed as follows:

$$Q = v_{10} \cdot Q_v + v_{11} \cdot Q_a + v_{12} \cdot Q_v \cdot Q_a + v_{13} \quad (8)$$

where $Q_a$ represents the audio quality, which can be further categorized into $Q_{as}$ and $Q_{am}$ for speech and music, respectively. Based on the results from Session 3 of Experiment 2 and findings from studies on audio and video playback speed, the model coefficients were determined separately for speech and music. In this audiovisual model, the coefficients $v_{10}$, $v_{11}$, and $v_{12}$ represent the relative contributions of video quality, audio quality, and their interaction, respectively, while $v_{13}$ serves as an offset term. These coefficients allow the model to account for the different integration patterns of audio and video quality in speech-video and music-video content. When the audio content is speech, the coefficients are as follows: $v_{10}$ = -0.281, $v_{11}$ = 0.197, $v_{12}$ = 0.186, and $v_{13}$ = 1.144. When the audio is music, the coefficients are as follows: $v_{10}$ = 0.791, $v_{11}$ = 0.327, $v_{12}$ = -0.023, and $v_{13}$ = -0.204.

TABLE VII
PERFORMANCE VALIDATION

| Types | Model | PCC | SROCC | RMSE |
|---|---|---|---|---|
| Video | CAPQ-FAST | 0.951 | 0.944 | 0.356 |
| | Baseline1 | 0.828 | 0.872 | 0.725 |
| Speech | CAPQ-FAST | 0.969 | 0.953 | 0.246 |
| | Baseline2 | 0.783 | 0.855 | 0.638 |
| Music | CAPQ-FAST | 0.957 | 0.931 | 0.269 |
| | Baseline3 | 0.614 | 0.584 | 0.718 |
| Video-Speech | CAPQ-FAST | 0.976 | 0.966 | 0.190 |
| | Baseline4 | 0.869 | 0.841 | 0.425 |
| Video-Music | CAPQ-FAST | 0.968 | 0.961 | 0.205 |
| | Baseline4 | 0.858 | 0.828 | 0.593 |

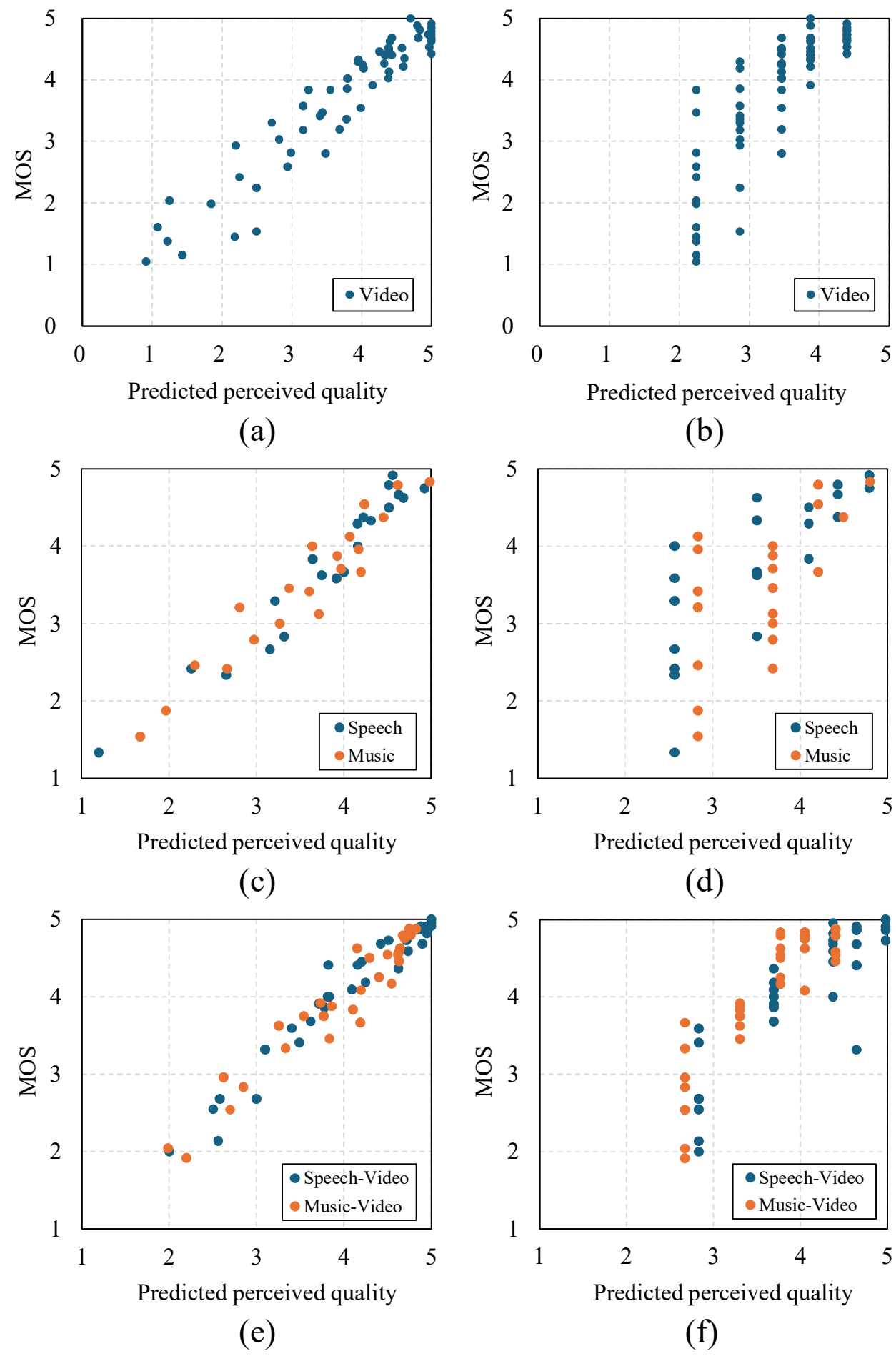

Fig. 7. Scatter plots of the predicted perceived quality and MOS. (a) CAPQ-FAST (video), (b) Baseline 1 model, (c) CAPQ-FAST (audio), (d) Baseline 2 model, (e) CAPQ-FAST (audiovisual), (f) Baseline 3 model.

## V. PERFORMANCE EVALUATION

This section mainly verifies the performance of the proposed model, which mainly includes two parts: (1) validating the prediction performance of the proposed perceived quality assessment model, and (2) examining the contribution of the proposed model in predicting users' willingness to adopt faster playback in real-world applications.

### *A. Performance Validation of the Proposed Model*

The performance of the proposed model is verified by comparing the predicted perceived quality with MOS using the data from Session 1 in Experiment 3. As far as we know, there are currently no objective models specifically designed for assessing the perceived quality of fast playback in video, audio, or audiovisual content. Therefore, several baseline models are selected for comparison to evaluate the effectiveness of the proposed quality assessment model under faster playback conditions.

For video faster playback, the baseline (Baseline 1) adopts the exponential form in (1), where the parameters $\alpha$, β, and $\delta$ are obtained by least-squares fitting on all data from Fig. 4 without distinguishing content, resulting in values of 6.345, 0.898, and 1.810, respectively. For audio faster playback, the exponential forms in (4) and (6) serve as the baseline models for speech and music (denoted as Baseline 2 and Baseline 3), respectively, without accounting for content characteristics such as speaking rate or musical tempo. The parameters $\lambda$ and $\mu$ of Baseline 2 model are fitted using data from Fig. 6(a) and are equal to 6.553 and 0.313, respectively. The parameters $\rho$ and $\gamma$ of Baseline 3, estimated using data from Fig. 6(b), are equal to 6.252 and 0.264, respectively. For audiovisual faster playback, the baseline model (Baseline 4) follows the structure presented in (8), where the video quality $Q_v$ is predicted using the Baseline 2 model and the audio quality $Q_a$ is estimated using the Baseline 3 models. Collectively, all these baseline models disregard content-specific characteristics of both video and audio.

To validate the model's performance, three evaluation metrics recommended by VQEG were adopted [18], namely Pearson Correlation Coefficient (PCC), Spearman Rank Order Correlation Coefficient (SROCC), and Root Mean Squared Error (RMSE). In general, lower RMSE values and higher PCC and SROCC values indicate better performance. Table VII summarizes the evaluation results of video, audio, and audiovisual on these metrics. Additionally, Fig. 7 presents scatter plots comparing the MOS values with the predicted perceived quality, providing an intuitive visual representation of the model's performance. The results show that the proposed model outperforms baseline models, with predictions closely aligning with subjective user ratings.

In addition, following the approach described in [14], an F-test was conducted on the residuals between the MOS scores and the predicted perceived quality to evaluate the significance of performance differences between the baseline models and the proposed model. This analysis was performed separately for video, audio, and audiovisual. The results of the F-test showed that all p-values (sig.) were below 0.05, indicating statistically significant differences between the proposed model and the baseline models. Moreover, the findings confirm that the proposed model offers a substantial improvement in prediction accuracy.

### *B. Application Validation for Coarse-Grained Playback Intention Interpretation*

To examine the usefulness of perceived quality as an auxiliary cue for interpreting faster playback behavior, this section analyzes whether users' reported playback intentions differ across perceived quality levels. According to the results from Session 2 of Experiment 3, Table VIII presents the proportion of user preferences (EV or SS) at each playback speed under different levels of perceived quality. Specifically, the baseline represents the proportion of user preferences at each speed without considering perceived quality. For comparison, Fig. 8 shows the percentage of users choosing EV across different perceived quality levels.

TABLE VIII
PLAYBACK INTENTION VALIDATION

| Perceived Quality | User Intention | Playback Speed | | | |
|---|---|---|---|---|---|
| | | 1.25x | 1.5x | 2x | 3x |
| PQ≥4 | EV | 89.3% | 82.3% | 63.3% | 45.5% |
| | SS | 10.7% | 17.7% | 36.7% | 54.5% |
| 3≤PQ<4 | EV | 89.5% | 80.0% | 68.9% | 42.9% |
| | SS | 10.5% | 20.0% | 31.1% | 57.1% |
| PQ<3 | EV | - | 0% | 12.0% | 7.4% |
| | SS | - | 100% | 88.0% | 92.6% |
| Baseline | EV | 87.5% | 77.1% | 53.0% | 27.1% |
| | SS | 12.5% | 22.9% | 47.0% | 72.9% |

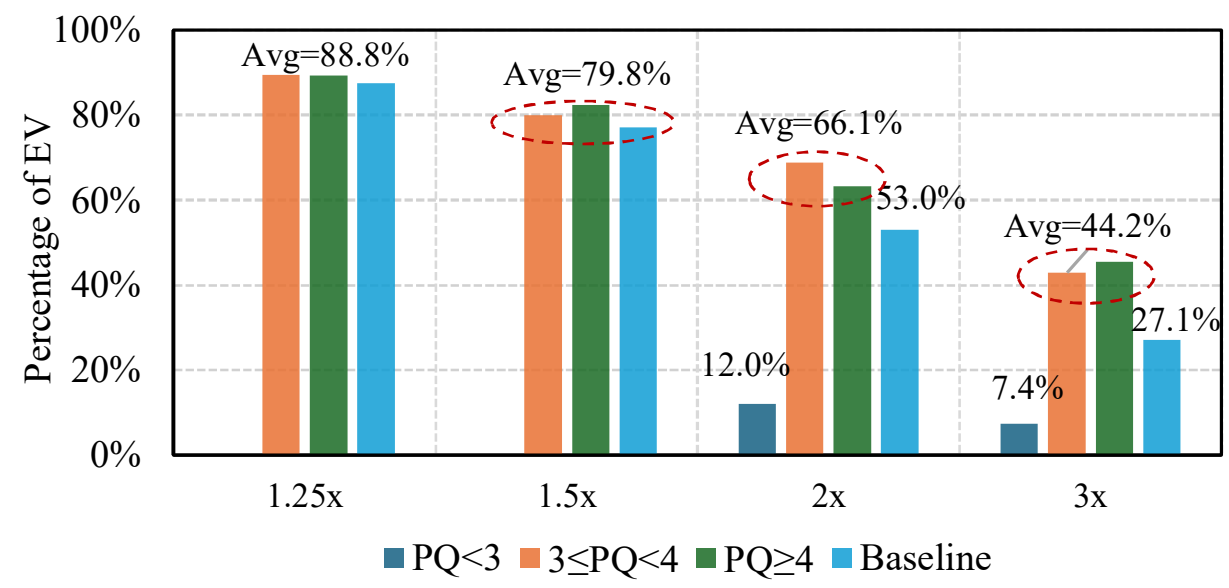

Fig. 8. percentage of users choosing EV under different perceived quality levels.

The results show that when playback speed does not exceed 1.5x, perceived quality remains within an acceptable range (greater than 3), and the predictions from the baseline and quality-informed models are relatively close. In this case, most users tend to choose EV, with an average of approximately 88.8% at 1.25x speed and 79.8% at 1.5x speed. However, at higher playback speeds such as 2x, there is a notable deviation between the EV selection rates under different levels of perceived quality and the baseline rate (53.0%). When perceived quality exceeds 3, the average EV selection rate increases to 66.1%, resulting in a deviation of 13.1% with the baseline. In contrast, when perceived quality falls below 3, the EV selection rate drops significantly, showing a 41% deviation from the baseline. This indicates that at 2x playback speed, users tend to prefer EV (avg. 66.1%) when the quality is acceptable, but favor SS when it is lower. This difference in preference can be better judged using perceived quality.

A similar trend is observed at 3x playback speed. When perceived quality exceeds 3, the deviation between the percentage of users choosing EV (avg. 44.2%) and the baseline (27.1%) is 17.1%. When the perceived quality is less than 3 points, the deviation increases to 19.7%. These findings indicate that user preferences at higher playback speeds are highly sensitive to perceived quality. Thus, incorporating perceived quality provides a useful auxiliary basis for interpreting users' faster playback intentions under different quality conditions. Nevertheless, such interpretation is coarse-grained and should be combined with semantic understanding, user profiles, and viewing history for more comprehensive intention modeling.

## VI. CONCLUSION

This study investigates the relationship between playback speed and perceived quality for different content modalities, including video, audio, and audiovisual. The findings confirm that perceived quality is not solely determined by playback speed, but is also closely linked to content characteristics. For video with low motion complexity and audio with slower speech or musical tempo, users tend to tolerate higher playback speeds without significant quality degradation. In contrast, content with fast-moving visuals or rapid speech and rhythm results in a lower tolerance for speed increases. Based on these confirmed observations with the associated quantitative findings, a content-aware perceived quality assessment model (CAPQ-FAST) was proposed to objectively evaluate perceived quality under various playback speeds for different media types. Experimental results demonstrate the strong predictive performance of the model, with potential applications in adaptive playback control, personalized speed recommendations, and improved user experience evaluation in online education, entertainment, and short-video platforms. Furthermore, the predicted quality scores effectively support the inference of user playback behavior intentions, providing a scientific basis for optimizing playback strategies and content recommendation systems in audiovisual platforms.

Although the current study covers video, speech, music, and audiovisual content, the dataset scale, content diversity, and participant diversity remain limited. Moreover, the current TI-, WPM-, and BPM-based temporal features mainly capture temporal dynamics and information density, rather than semantic importance, narrative relevance, or user interest. Future work will expand the dataset, include more diverse participants, and explore semantic-aware visual, audio, or multimodal representations to improve model generalizability and user behavior understanding.